\documentclass[aps,prc,groupedaddress,showpacs,twocolumn,floatfix]{revtex4}

\usepackage{graphicx}
\usepackage[dvips]{color}
\usepackage{colordvi}

\def\gtorder{\mathrel{\raise.3ex\hbox{$>$}\mkern-14mu
 \lower0.6ex\hbox{$\sim$}}}
\def\ltorder{\mathrel{\raise.3ex\hbox{$<$}\mkern-14mu
 \lower0.6ex\hbox{$\sim$}}}

\def\beq{\begin{equation}}
\def\eeq{\end{equation}}
\def\ba{\begin{eqnarray*}}
\def\ea{\end{eqnarray*}}

\begin{document}

\title{Correlated nucleons in configuration space }

\author{Herbert M\"{u}ther}
\affiliation{Institut f\"{u}r Theoretische Physik, Universit\"{a}t T\"{u}bingen,
T\"{u}bingen, Germany}

\author{Ingo Sick}
\affiliation{Dept. f\"{u}r Physik und Astronomie, Universit\"{a}t Basel, 
Basel, Switzerland}

\date{\today}

\begin{abstract}

Several recent studies have dealt with the effects of short range correlations
on the momentum distribution of nucleons in nuclei. Here we investigate the
correlation effects on the density and spectral distribution in coordinate
space. A combination of experimental data and spectral functions calculated
from realistic N-N interactions allows us to resolve a recently uncovered
discrepancy with  occupation of quasi-particle states derived from (e,e'p)
data.

\end{abstract}


\maketitle

\section{Introduction}
Much of the understanding of atomic nuclei is based on the assumption that 
nucleons  move,
independently from each other, in the average potential created by the
interaction with all other nucleons.  A more fundamental approach to the 
understanding of nuclei has to start from the
underlying nucleon-nucleon (N-N) interaction. Realistic models of the
N-N interaction exhibit a strongly repulsive central
interaction at small inter-nucleon distances and a strong tensor
component. These features lead to properties of nuclear wave functions that are
beyond what is describable in terms of  Independent Particle (IP) motion. In 
particular, strong  short-range correlations (SRC) are expected to occur.

The effects of the short-range correlations are known for systems where an
accurate solution of the Schr\"odinger equation for a realistic N-N interaction
can be obtained \cite{Pandharipande97}. Very light nuclei (today up to A$\leq$10) and infinite nuclear
matter are amongst the systems where this is feasible 
\cite{Pieper01,Benhar89,Muther00}. The corresponding
calculations show that in a microscopic description of nuclear systems 
the short-range and tensor parts of
the N-N interaction have a very important, not to say dominating, influence
without which not even nuclear binding can be explained. 

  Due to these short-range correlations the momentum distributions of 
nucleons acquire a tail extending to very high momenta $k$ and 
part of the strength, located in IP descriptions at low excitation energy 
$E$, is moved to very high excitation energies.

In the past, most experimental investigations were confined to rather low
momenta and energies, {\em i.e.} to the region where the strength is dominated
(but not entirely given) by the IP properties. In this region, the consequences
of short-range correlations are indicated primarily by a {\em depopulation} of states in
comparison to the predictions of IP models (including the long-range correlations
which can be described by configuration mixing). According to the 
calculations mentioned above, a depopulation of the order of 20\% is expected.

>From the experimental information available up to now, the  
depopulation of IP strength at low $k,E$ is established \cite{Pandharipande97}
 (for a caveat see  below). Much less is known from
a {\em direct} measurement of the strength of the spectral function $S(k,E)$ at
large $k$ and $E$. A recent (e,e'p) experiment, performed at high momentum
transfer $q$ in parallel kinematics by Rohe {\em et al.}, provides the first direct measurement 
\cite{Rohe04,Rohes04a}. 

This correlated strength has always been discussed in $k,E$-space where (part
of) it can be separated from the IP strength. In this Rapid Communication, we take
an orthogonal look at the correlated strength and discuss it in coordinate
space (r-space). 
We address this question from both the theory and experiment side.

The trigger for this study lies in  difficulties experienced in the past in
interpreting data in terms of IP models. For example,  fits with IP 
wavefunctions
of the nuclear charge density often yielded form factors (of IP dominated 
transitions) with
incorrect $q$-dependence. Fits with IP wave functions also have difficulties to
reproduce the total densities in the nuclear interior.  The origin of these
difficulties: total densities have contributions from correlated nucleons that
do not appear in observables dominated by quasi-particle  properties. The
correlated nucleons presumably have a different radial distribution. 

The goal of this paper is to derive quasi-particle (QP) and correlated
distributions in r-space using Green's-function theory. We compare the results
to the correlated density in r-space which we obtain from the difference of
the  density --- known from elastic electron scattering --- and the QP
contributions known from (e,e'p) reactions.    

As a side-product, this study also sheds light on a recently uncovered problem
with QP occupation numbers derived from (e,e'p) experiments with low and high
$q$, respectively \cite{Lapikas00}. 

\section{Single-particle spectral function}
The evaluation of the single-particle spectral functions for $^{12}C$, the
nucleus we use for our study,  has been
performed within the framework of the Green's function method \cite{Dickhoff04}
using the techniques described in \cite{Muether95,Polls95,Polls97}. The nucleon
self-energy $\Sigma_{lj}(p_m,p_n,E)$ is determined in a discrete basis of
Bessel functions $\phi_{p_mlj}(r)$ with appropriate boundary conditions at the
surface of a spherical box with radius $R_{\mbox{box}}$. These basis states are
identified by the angular momentum quantum numbers $l$ and $j$ and a radial
quantum   number ($p_m$, $p_n$) referring to the momentum. For a box radius
$R_{\mbox{box}}$     of typically 20 fm, it turns out to be sufficient to
include around 60 basis states for each partial wave. 

The self-energy contains a Hartree-Fock contribution 
$\Sigma_{lj}^{\mbox{\scriptsize HF}}$
 calculated in terms of a nuclear matter $G$-matrix plus complex correction 
terms, $\Delta \Sigma_{lj}$, which account for the inclusion of two-particle one-hole
and two-hole one-particle contributions. These correction terms are calculated
directly for the finite nucleus $^{12}C$, describing the intermediate particle
states by plane waves orthogonalized with respect to the occupied hole states.
This is a good approximation to describe the effects of SRC, however, it
tends to underestimate the spectral strength due to long-range correlations at
missing energies slightly above the two-hole one-particle threshold.

The single-particle Green's function is determined from this self-energy
$\Sigma_{lj}^{\mbox{\scriptsize HF}}+\Delta \Sigma_{lj}$ by solving the Dyson equation in the
box basis described above. From the imaginary part of this Green's function one
can calculate the spectral function in this basis \cite{Polls97} or determine it
in configuration space by the transformation
\ba
S_{lj}(r,r';E) = \sum_{m,n} \phi_{p_mlj}(r)
S_{lj}(p_m,p_n;E)\phi^*_{p_nlj}(r')\,,\label{eq:spectr1}
\ea
using the Bessel functions $\phi_{p_mlj}(r)$ described above. The spectral
function can be split into the QP contribution $S^{QP}_{lj}$, which only occurs in 
the $s_{1/2}$ and $p_{3/2}$ partial waves, and in the continuum contribution,
$S^{cont}_{lj}$, which originates from the imaginary components in the
self-energy. This leads to the single-particle density
\ba
\rho(r) & = & \sum_{lj} S^{QP}_{lj}(r,r) +
\sum_{lj}\int_{\varepsilon_{2h1p}}^{\infty} dE\,
S^{cont}_{lj}(r,r;E)\nonumber\\
& = & \rho_{QP}(r) + \rho_{corr}(r)\,,\label{eq:densit}
\ea
where the integration over missing energies $E$ starts at the threshold of
two-hole one-particle configurations. We have assigned the label $corr$ to the
part of the single-particle density, which originates from the continuum part
of the spectral function to indicate that this correlated density is absent in
the IP model.

\begin{figure}[htb]
\includegraphics[height=7.5cm,width=8.5cm,angle=0]{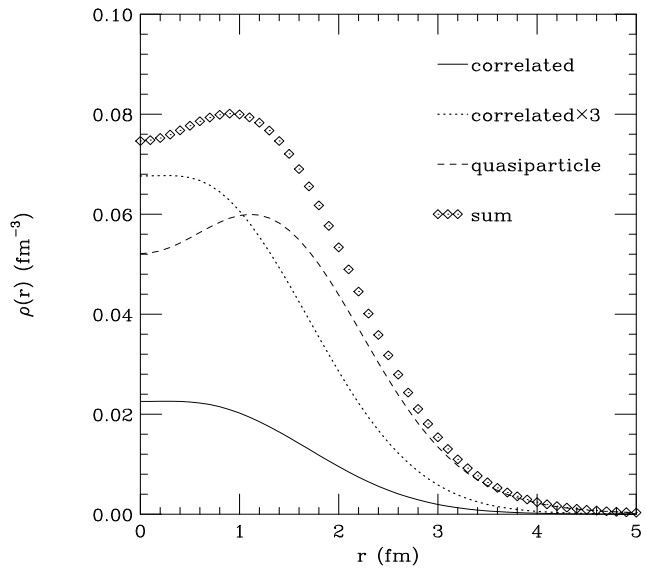}
\caption{Comparison of QP and correlated densities from theory.   
\label{rhomuth}}
\end{figure}

Results for these contributions to the point density of protons in $^{12}C$ are
displayed in Fig.~\ref{rhomuth}. The calculation of these densities have been
performed using the CD-Bonn potential for the N-N interaction \cite{CDBonn}.   A
fraction of the proton density, which accounts for around 5 protons, is
described by the QP part and  the rest is covered by the correlated density. 

In order to allow for a better comparison of the radial shape of the 
density contributions,  Fig.~\ref{rhomuth} also contains the correlated density
$\rho_{corr}(r)$  multiplied by a factor of 3. The comparison shows quite
clearly that the correlated density is located much more in the center of the
nucleus than the QP contribution. 

The correlated single-particle density is distributed over partial waves $lj$
 including those which are unoccupied in the IP model.
A large fraction of the correlated strength, however, is contained in partial
waves with $l=0$ (around 31\%) and $l=1$ (around 37\%). One also should note
that, contrary to what one naively could  expect, the strength in the higher 
$l$ states does
not contribute at large $r$; the corresponding large values of $E$  pull the
radial wave functions to lower $r$.  

\section{Densities from $(e,e)$} 
For $^{12}C$ an extensive set of elastic
electron scattering data is available \cite{Sick70,Cardman80,Fey73,Jansen72}
covering the range of momentum transfer between 0.13 and 3.7 fm$^{-1}$. The
Carbon $rms$-radius is precisely known from a  $\mu$-X-ray experiment
\cite{Ruckstuhl84}.

These data have been used to determine a model-independent charge density using
the SOG-approach  of \cite{Sick74}. The procedure employed has been described in
\cite{Sick82}.  This yields the charge density as a function of $r$, together 
with
an error bar that covers both the statistical and systematic uncertainties of
the data, as well as the model-error. 

In order to obtain the point density, we have unfolded the effect of the finite
size of the proton and neutron charge density. This has been done by
parameterizing the point density, folding it with $\rho_p(r) + \rho_n(r)$ and 
fitting the resulting density to the charge density as determined above.
The contribution of the electro-magnetic  spin-orbit term to the charge density turned out to
be negligible. The resulting folded density agrees within $\sim$1\% with the one
given in \cite{DeVries87}.  

The resulting point density is shown in figs.~\ref{rosum1},\ref{rosum2}. The error bars in
general are too small to be seen.

\section{QP orbits from $(e,e'p)$}
For Carbon, quite an extensive set of (e,e'p) data is available 
\cite{Amaldi67,Nakamura76,Mougey76,Bernheim82,Steen88,Blomqvist95b,Makins94,Dutta03};
 a compilation
is discussed in \cite{Lapikas00}. Part of this data has been taken at low $q$
with the goal to determine the 1p and 1s quasi-particle momentum distributions
and occupation numbers. Some data have been taken at large $q$ mainly in
connection with the determination of nuclear transparencies for high-energy
protons.

The low-$q$ data,  taken with good energy- and momentum resolution mainly 
at NIKHEF and Saclay, have been analyzed in the framework of DWBA using 
optical potentials known from
proton-Carbon scattering. The QP  radial wave functions have been parameterized
using Woods-Saxon (WS) potentials. Lapikas {\em et al.} \cite{Lapikas00} have made a 
coherent analysis of the entire data set. The occupation of the QP orbits,
obtained by summing the experimental spectroscopic factors,  
turns out to be rather low in comparison to what is known for other nuclei
\cite{Pandharipande97}; the
summed 1p plus 1s strength amounts to 3.4 protons only (56\% occupation). 

The high-$q$ data, determined in part  with moderate energy- and momentum resolution, were
taken mainly at SLAC and JLAB. The data were summed over a large region of
initial momentum $k$ and removal energy $E$, and fitted using WS radial wave
functions and theoretical transparencies. When using the most reliable
transparencies, confirmed by other experiments, Lapikas {\em et al.} found a much
higher occupancy of the QP orbits, 5.0-5.6 protons ($\sim$87\% occupation).    

This discrepancy --- which is very embarrassing to the practitioners of (e,e'p) as
it sheds serious doubts on the quantitative interpretation of (e,e'p) data --- 
obviously needs to
be better understood, and is discussed in more detail below.  This difference
also has led to speculations about $q$-dependent QP-occupations, for which we
see no physical basis.

A partial reason for the difference between the low-$q$ and high-$q$ results is
immediately clear: The low-$q$ data cover the region of missing momenta of
typically $<$180MeV/c and missing energy $E$ $<$50MeV, the high-$q$ data 
extend to 300MeV/c. 
The high-$q$ data
also cover a larger range in missing energy, they are integrated up to typically
80MeV. In this larger $k,E$-range, there is not only QP strength, but also a
fraction of the correlated strength is integrated over. Before making a valid
comparison, this correlated strength needs to be removed. 

In order to correct for this effect, we start from the high-$q$ (e,e'p) data
taken in a recent JLAB experiment \cite{Rohe04} in quasi-elastic kinematics, 
which minimize final state interactions FSI and meson exchange current
contributions MEC. This experiment yields, in agreement with the previous
JLAB and SLAC experiments, 5.2 protons in the integration region E$<$80MeV,
k$<$300MeV/c. This number we correct for the continuum contribution using
the calculated spectral function discussed above.  
With this correction the discrepancy between the low-$q$ and high-$q$ results
is significantly reduced; the QP occupations now are 3.4 {\em vs} 4.5 protons
for the low-$q$/high-$q$ data, respectively.
Given the uncertainties of these numbers --- believed to perhaps 10\% --- there
still is a worrisome  incompatibility. 

In order to proceed, we have to choose. We have decided to use  the
occupation coming from the high-$q$ measurements, as we judge the interpretation
of these data to be more safe. The low-$q$ data suffer from uncertainties in
the treatment of the final state interaction. Due to the low
energies of the outgoing proton (70MeV for the NIKHEF data \cite{Steen88})
coupled channel effects not treated in the usual DWBA analysis should be
relevant. Van der
Steenhoven {\em et al} \cite{Steen88,Steen87c} have shown that inclusion of
these effects would increase, for the rather soft nucleus $^{12}C$,  
the QP occupation by $\sim$20\%. For the kinematics
of the low-$q$ experiments, the calculations of Boffi {\em et al.}
 \cite{Boffi91,Boffi93} also predict significant MEC effects that would lead to
 a further  increase of the QP occupation.  
   
The  value for the QP occupation deduced from the high-$q$ measurement is
also compatible with the correlated strength measured directly in the recent
experiment by Rohe {\em et al.} \cite{Rohe04}. 
 This measurement agrees with theoretical predictions for
the correlated strength of $\sim$20\%. The summed QP strength from the low-$q$
(e,e'p) data (which thus includes the fragmentation due to long-range
correlations), on the other hand, would correspond to $>$40\% correlated 
strength, {\em i.e.} be unrealistically high.
 Furthermore, as we will see
below, our choice of the QP occupation is confirmed by a consistency check in
our analysis.

Before proceeding to the calculation of the QP density, one more effect of
correlations must be removed from the results of Lapikas {\em et al}. While the
p-strength is located in discrete states where no ambiguity occurs, the
s-strength is located in the continuum between 20 and $\sim$50 MeV removal
energy. In this region, also the correlated strength contributes, and affects
the {\em shape} of the fitted WS momentum distribution. The correlated strength has 
a momentum
distribution that falls much more slowly with increasing $k$ than the 1s QP
strength; when fitting the sum with a WS parameterization, the resulting WS
momentum distribution extends somewhat too far in momentum, {\em i.e.} it would
have too small a radial extension in $r$-space. 

We have used the theoretical 1s QP and correlated strength in the region used in
\cite{Lapikas00} for the determination  of the 1s momentum distribution to
calculate a correction to the  WS parameterization fitted to the sum. 
We find that the extent of
the wave function in k-space needs to be reduced by 11\%. This modified
WS-shape has been
confirmed \cite{Rohe04b} by an independent analysis employing the recent JLAB 
data of \cite{Rohe04} and the correlated spectral function of \cite{Benhar89}. 

With these QP radial wave functions, and occupations renormalized to the one
derived above from the high-$q$ experiments, we can compute the QP density in
$r$-space. As the radial wave functions fitted by Lapikas {\em et al.} refer to
relative coordinate between the proton and the CM of the (A--1) system, we need
to rescale the radial coordinate by a factor 11/12, with the corresponding
adjustment in height to conserve the normalization. 

\begin{figure}[htb]
\includegraphics[height=7.5cm,width=8.5cm,angle=0]{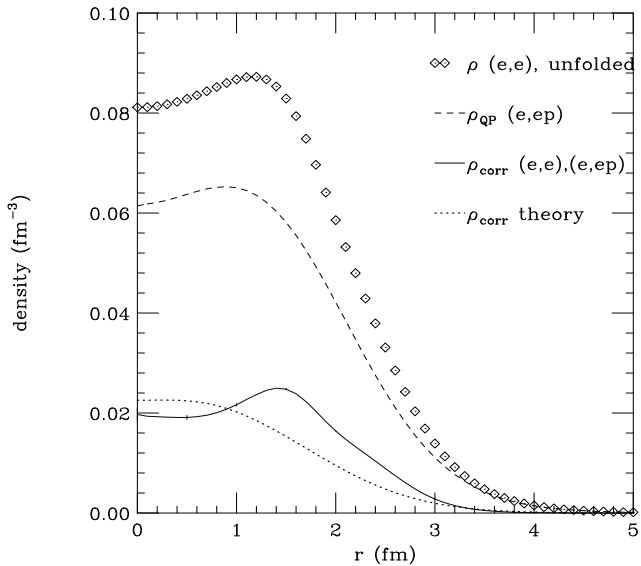}
\caption{Comparison of QP and correlated densities.
\label{rosum1}}
\end{figure}

\begin{figure}[htb]
\includegraphics[height=9.5cm,width=8.5cm,angle=0]{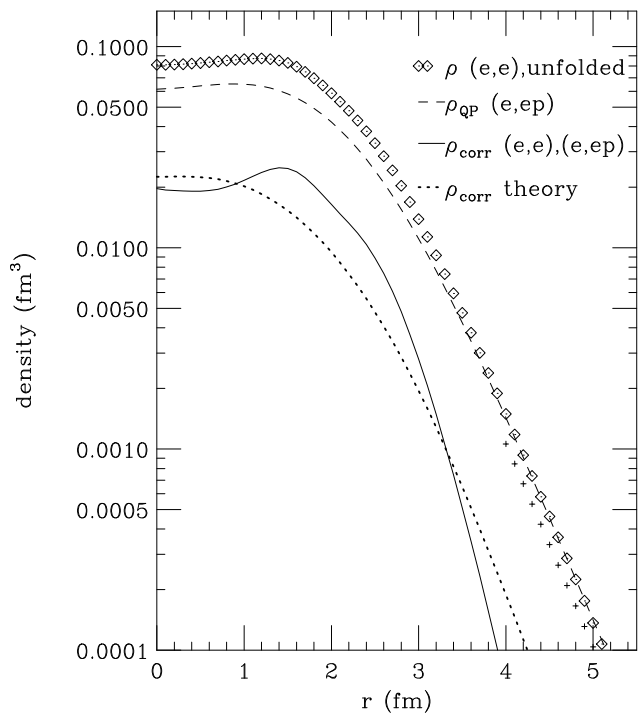}
\caption{Comparison of QP and correlated densities. The crosses indicate the
tail of the density obtained when using the low-$q$ QP occupation.  
\label{rosum2}}
\end{figure}

The resulting QP density is shown in figs.~\ref{rosum1},\ref{rosum2}. 
The difference
of the $^{12}C$ point density and this QP density yields the  correlated density
also shown in Figs.~\ref{rosum1},\ref{rosum2}. The 'error bar' on the correlated
density is not straightforward to calculate due to the various adjustments that
had to made in the analysis; $\pm$20\% for $r<3fm$ is probably a realistic estimate. 

\section{Comparison of QP and correlated densities}
The first observation one can make from Figs.~\ref{rosum1},\ref{rosum2} concerns the
fact that like in the theoretical analysis of Fig.~\ref{rhomuth} also the 
correlated density deduced from experimental data is significantly more concentrated towards the
nuclear interior than the QP density. One also observes that the correlated density in the
nuclear interior gives a very significant contribution, of order 30\%, of the
central density,  {\em i.e.} larger than one could have expected from the 
number of 20\% or so  of correlated nucleons.  This explains why attempts to explain the total densities in
terms of QP orbitals cannot be very successful.

Figure \ref{rosum2} shows another important feature: The experimental QP density
at large $r$ agrees perfectly with the point density measured via elastic
electron scattering. Such an  agreement  should occur, as at large $r$ ---
outside the range of the nuclear potential --- the density is entirely given by
the tail of the least-bound QP orbit, the 1p$_{3/2}$ state in $^{12}C$. 
More deeply bound states, or correlated nucleons with large removal energy,
cannot contribute.

The good agreement between QP density and point density at large $r$ also
confirms the correctness of our choice of QP occupations. Had we used the
occupation derived from the low-$q$ experiments and, as above, the shape of R(r)
determined in \cite{Lapikas00} from the fit to the world (e,e'p) data, we would have obtained the tail
indicated in fig.~\ref{rosum2} by the crosses. These are obviously significantly
too low. This comparison thus provides an {\em a posteriori} justification of our
procedure.

In figs.~\ref{rosum1},\ref{rosum2} we also show the correlated density obtained by
theory. Considering the above-mentioned uncertainty of the experimental result,
we consider the agreement between theory and experiment as a good one. The size
of the correlated contribution in the nuclear interior is very similar, the
rapid fall-off of the correlated density at large $r$ also agrees within our
estimated uncertainty. The
correlated density deduced from experimental data seems to contain more spectral
strength in partial waves with $l>0$ than the theoretical one.

The large  contribution of the correlated density in the nuclear interior shows
that the neglect of this correlated contribution in the standard IP calculations
({\em e.g.} all the shell-model descriptions) is not justified.     

\section{Conclusions}
Starting from (e,e'p) data we have constructed the QP density for $^{12}C$
in coordinate space. The difference to the total density, obtained from elastic
electron scattering, provides the density distribution of the correlated
nucleons. We find that it is significantly more concentrated towards the nuclear
interior. We also find good agreement with the theoretical calculation of the
correlated density distribution.

The large contribution of the  density related to short-range N-N correlations, 
$\sim$30\% in the nuclear
interior, together with the fact that the {\em shape} of the correlated density
differs strongly from the QP density, explains the poor performance of QP wave
functions in explaining many observables. Due to the shape difference, the
shortcoming of the neglect of the correlated contribution also cannot
satisfactorily be 'compensated' by using effective quantities like effective
charges, {\em etc}. 

As a side product, our analysis provides a solution to the puzzle raised in
\cite{Lapikas00}, the pronounced disagreement between QP occupations derived
from the low-$q$ and high-$q$ (e,e'p) experiments. We find that only the
high-$q$ occupation (suitably corrected for the correlated contribution
not considered  in \cite{Lapikas00}) is compatible with the independent 
information 
from elastic electron scattering.

\begin{acknowledgments}
We would like to thank Louk Lapikas for providing the QP wave functions,  Kai
Hencken and  Dirk Trautmann  for elucidation of CM problems and Daniela
Rohe for an independent determination of the 1s QP momentum distribution. 
This work has  been supported by the DFG
(Graduiertenkolleg Basel-T\"ubingen) and the Schweizerische Nationalfonds.

\end{acknowledgments}

  \end{document}